\documentclass[12pt,preprint]{aastex}

\lefthead{Cunha et al.}
\righthead{Fluorine Abundances in the Milky Way Bulge}

\begin{document}

\title{Fluorine Abundances in the Milky Way Bulge}

\author{Katia Cunha\altaffilmark{1}, \& Verne V. Smith}
\affil{National Optical Astronomy Observatory,
Casilla 603, La Serena, Chile; kcunha@noao.edu; vsmith@noao.edu}
                                                                                
\author{Brad K. Gibson}
\affil{University of Central Lancashire,
Preston PR1 2HE, UK; bkgibson@uclan.ac.uk}

\altaffiltext{1}{On leave from
Observat\'orio Nacional; Rio de Janeiro, Brazil}

\begin{abstract}
Fluorine ($^{19}$F) abundances are derived in a sample of 6 bulge red giants
in Baade's Window.  These giants span a factor of 10 in metallicity
and this is the first study to define the behavior of $^{19}$F with 
metallicity in the bulge. The bulge results show an increase in F/O with increasing
oxygen. This trend overlaps what is found in the disk at comparable
metallicities, with the most oxygen-rich bulge target extending the disk trend. 
The increase in F/O in the disk  arises from $^{19}$F synthesis in both
asymptotic giant branch (AGB) stars and metal-rich Wolf-Rayet (WR) stars through stellar winds.  
The lack of an s-process enhancement in the most fluorine-rich bulge giant in this study,
suggests that WR stars represented a larger contribution than AGB stars to $^{19}$F 
production in the bulge when compared to the disk. If this result for fluorine is
combined with the previously published overall decline in the O/Mg abundance ratios in metal-rich 
bulge stars, it suggests that WR winds played a role in shaping chemical evolution in the bulge.
One star in this study exhibits a very low value of F/O while having a large O-abundance; this
chemical mixture can be understood if this star formed from gas that was
enriched by metal-poor core-collapse supernovae and may indicate that
chemical evolution in the bulge was inhomogeneous.
\end{abstract}

\keywords{stars: abundances; Galaxy: abundances; Galaxy: bulge}

\section{INTRODUCTION}

Understanding how chemical evolution has proceeded in the Galactic bulge can
provide clues for models of bulge formation and evolution.  It is not known, for example,
whether the Milky Way bulge was formed rapidly in a single collapse
or via secular dynamical evolution driven by the disk. 
Certain elemental abundance ratios can be used to infer
timescales for chemical enrichment within a particular stellar population.  
The most studied of these ratios involves comparing the abundances of the so-called 
$\alpha$-elements (such as O, Mg, or Ca), 
which are produced via massive-star core-collapse supernovae of type II (SNII),
to abundances of iron, which is produced in SN Ia.
Probing elemental species that are
created in other astrophysical sites, such as AGB stars 
or WR stars, can add further constraints to bulge formation scenarios.

The first study to provide chemical abundance distributions of several elements
in a sample of bulge red giants was 
McWilliam \& Rich (1994).
It is only within 
the last few years that additional abundance studies have appeared, all of which
rely on the 8-10m class telescopes.   These recent studies include, in the optical,
Zoccali et al. (2006); Fulbright et al. (2006, 2007); Lecureur et al. (2007); McWilliam 
et al. (2007) 
and, in the infrared,  Rich \& Origlia (2005);
Cunha \& Smith (2006); Rich et al. (2007) and Melendez et al. (2008).
Although a relatively large number of bulge targets have been studied so far,
the abundance patterns of the Galactic bulge population continue to be probed
in increasing detail.

One element that can add new insight into the nature of chemical evolution
in the bulge is fluorine.  Understanding the origins of this light element has advanced
considerably in recent years, based upon $^{19}$F abundances derived from infrared
vibration-rotation lines of HF (Jorissen et al. 1992; Cunha et al. 2003; Cunha \& Smith
2005; Smith et al. 2005).  Renda et al. (2004) use the observed abundances to model the
Galactic chemical evolution of fluorine, with its synthesis occurring primarily
in three different astrophysical sites:
in AGB stars as a result of He-burning (Goriely et al. 1989; Forestini et al. 1992; Jorissen et al. 1992), 
in SN II via neutrino nucleosynthesis (Woosley et al. 1990; Woosley \& Weaver 1995), and in WR stars 
as a result of He-burning and extensive stellar winds (Meynet \& Arnould 2000).
Renda et al. (2004) found that neutrino nucleosynthesis was the important source of
$^{19}$F in the early Galaxy (at low metallicity); however, the fluorine abundances
found in near-solar metallicity stars required significant contributions from both 
AGB stars and WR winds.

This paper concentrates on determining fluorine abundances in a sample of red giants 
of the Galactic bulge and  these results are combined with previously derived
abundances from other elements.   Observational evidence for fluorine production
in WR stars, compared to AGB stars or neutrino nucleosynthesis in SN II, is
discussed as well as the implication for the nature of chemical evolution in the bulge.

\section{OBSERVATIONS}

The target stars for this analysis of fluorine 
were taken from our previous 
infrared high-resolution spectroscopic study 
of Galactic bulge giants (Cunha \& Smith 2006). 
The sample, which is composed of 5 K- and 2 M-giants, is presented in Table 1.  Details
about the nature of these stars, all of which lie in Baade's Window, can be found in Cunha \&
Smith (2006).  
The spectra were observed in queue mode with the 8.1m Gemini South
telescope and the NOAO spectrograph Phoenix (Hinkle et al. 1998) at a resolution R$\sim$50,000; these
were centered at 23400\AA\ in order to include the HF 1-0 R9 line and covered a window of
$\sim$120\AA.
The K-giants in our sample were observed in May and July 2004; June and July 2005 (same spectra
were analyzed previously for Na in Cunha \& Smith 2006); 
while the two M-giant observations were taken more recently during one night in June 2007.
A description of the Phoenix observations and the reduction of the high-resolution spectra 
can be found in Cunha \& Smith (2006) and Smith et al. (2002). 

\section{Analysis}

All target stars were previously analyzed in the literature and had stellar parameters
and microturbulent velocities (Table 1) derived in Cunha \& Smith (2006). The effective
temperatures were obtained
using calibrations of infrared photometry (J-K and/or V-K colors) and extinction maps of
Stanek (1996). The surface gravities were derived from standard relations 
between stellar luminosity and mass as defined by isochrones corresponding to 10 Gyr by 
Girardi et al. (2000). The microturbulent velocities were estimated from measurements
of CO molecular lines which are also present in the observed Phoenix spectra in the K-band.
More detailed information on the derivation of the stellar parameters can be found
in Cunha \& Smith (2006).

The fluorine abundances are derived from the HF 1-0 R9 line at 23357 \AA.
The reliability of this line as an acurate abundance indicator has been verified in 
Cunha et al. (2003) from comparisons with other HF lines (which were analyzed in 
Jorissen et al. 1992).  Fluorine abundances were obtained from 
synthetic spectra computed with an updated version of the synthesis code MOOG 
(Sneden 1973) and adopting MARCS model atmospheres (Gustafsson et al. 1975). 
Figure 1 shows both synthetic and observed spectra 
for one sample star. The derived fluorine abundances are presented in Table 1 in the
nomenclature of A(x)=Log[N(x)/N(H)] + 12.0.

In addition to the $^{19}$F abundances in Table 1, values for A(Na) are also shown, with 5
of the Na abundances taken from Cunha \& Smith (2006).  They are presented here along
with the two new Na abundance results for 
BMB78 and BMB289; from the Na I line at 23379\AA. Oxygen abundances are also included for
completeness with abundances taken from Cunha \& Smith (2006).

\section{DISCUSSION}

The chemical evolution of the Galatic bulge has been modelled recently by Ballero et al. (2007), 
who focused on the contraints provided by recently published abundances of iron 
and $\alpha$-elements in bulge red-giants. Although at the moment such models do not predict 
the evolution of the element fluorine in particular, 
the behavior of the fluorine abundances derived in this study can be used to 
interpret some aspects of chemical evolution in the bulge
population.  

This interpretation begins with Figure 2, 
where the ratio of F/O (Log[N(F)/N(O)]) is plotted as a function of 
the oxygen abundance, A(O), and oxygen is used as a proxy for the overall 
metallicity. The five bulge $^{19}$F measurements are
shown as the red circles, with estimated errors
indicated.  All results to-date for Galactic field stars are also
plotted, with these abundances taken from Cunha et al. (2003), Cunha \&
Smith (2005), and  Cunha et al. (2008).  
The two populations shown in Figure 2 (the Galactic field and the bulge) both exhibit
generally increasing values of F/O as the O-abundance increases.  
Overall, the bulge giants overlap the trend set by the field stars, with the most
O-rich bulge star studied (IV-072) apparently defining a smooth extension of the
field-star trend to ever increasing oxygen abundances.
One bulge M-giant, BMB78, defies the general trend by having a
relatively low value of F/O given its high oxygen abundance.

The solid line in Figure 2 represents the predicted values of
$^{19}$F/$^{16}$O, as a function of metallicity, derived from the Woosley \& Weaver (1995) SN II yields,
convolved with a Salpeter mass function and an upper limit
of 40M$_{\odot}$; the $^{19}$F from these models is produced by neutrino nuclesosynthesis.
More recently, however, Heger et al. (2005) argue that $^{19}$F production via neutrino
nucleosynthesis should be lowered by about a factor of two, due to reduced
cross-sections, and the dashed line in Figure 2 is a shift of the Woosley \& Weaver (1995)
yields downward by 0.3 dex as a simple way of viewing these suggested revisions.
It is clear from the results in the figure that 
at the lowest metallicities, the observed values of F/O for field disk stars tend to approach
the values predicted by the yields in which $^{19}$F is synthesized via neutrino nucleosynthesis.   
The Sun and near-solar metallicity field stars, however, fall above the predicted F/O values
from neutrino nucleosynthesis and this difference points to significant contributions
to $^{19}$F production from WR and AGB stars, as suggested by Renda et al. (2004).

With four out of the five bulge stars containing larger ratios of fluorine to oxygen than can
be accommodated by neutrino nucleosynthesis alone, one is left with two possibilities
for $^{19}$F production at high metallicities, based upon the Renda et al. (2004) model: 
the AGB and WR stars.  Can one now attempt to distinguish between these two sites for 
$^{19}$F production in the bulge,  keeping in mind that there are no bulge-specific chemical evolution 
models for $^{19}$F?
Looking first at the AGB stars, Jorissen et al. (1992) pointed out that there is a positive 
correlation between F/O with the s-process abundances (their figure 12) and this correlation
was modelled by Goriely \& Molawi (2000) for neutron capture nucleosynthesis in AGB stars.
Both the model predictions and the observed correlation between fluorine and s-process abundances
would suggest that the most fluorine-rich star observed in the bulge, IV-072, should be 
heavily enriched in s-process elements at the level of [s/Fe] $\sim$ +1.5 dex, if the $^{19}$F 
resulted from AGB production. However, McWilliam \& Rich (1994) derived abudances for two
s-process elements in IV-072 and obtained [Y/Fe]=-0.02 dex and [La/Fe]=-0.04 dex; far below what
would expected from AGB models and observed correlations.  In addition, recent results for
heavy-element abundances in three metal-rich bulge dwarfs, whose brightnesses were increased 
during microlensing events, do not find s-process enrichments: 
[s/Fe] $\sim$ +0.12 dex (Zr, Ba and La from Cohen et al. 2008); -0.24 dex (Ba from Johnson et al. 2008) 
and -0.28 dex (Ba from Johnson et al. 2007).   

Given the apparent lack of s-process enriched
high-metallicity bulge stars, the best explanation for the large F/O value in IV-072 may be 
WR fluorine production.  Such a conclusion is reached by Renda et al. (2004) for the
metal-rich end of disk chemical evolution.  
We note, however, the cautionary points raised by Palacios et al. (2005) in regard to $^{19}$F production 
in WR stars; rotationally-induced mixing and mass-loss prescriptions can in fact lead to either an 
order-of-magnitude decrease in $^{19}$F production (for high-mass ($>$30-80 M$_{\odot}$) fast rotators at
solar-to-supersolar metallicities) or an order-of-magnitude increase in  $^{19}$F production (for lower 
mass ($<$30 M$_{\odot}$) fast rotators at supersolar metallicities). The issue of the $^{19}$F($\alpha$,p)$^{22}$Ne 
reaction rate uncertainty raised by the downwards revision proposed by Lugaro et al (2004), 
appears ameliorated by the recent work of Ugalde et al (2008), which is consistent with the canonical 
rate of Caughlan \& Fowler (1988).
The large $^{19}$F abundance in IV-072 may
require a relatively large amount of WR-wind material sculpting the chemical evolution
of the metal-rich bulge population.  This conclusion, based on fluorine, agrees with conclusions
that are based on the ratios of O to Mg in metal-rich bulge and
disk stars by McWilliam et al. (2007).

While 4 out of 5 bulge fluorine abundances follow an increase in F/O as
the stellar metallicity increases, the peculiar position of BMB78 in Figure 2 questions whether
bulge metallicity increased in a monotonic fashion.
This star is quite oxygen-rich yet has a low fluorine
abundance: its value of $^{19}$F/$^{16}$O is consistent with the yields predicted from
neutrino nucleosynthesis only. The low value of F/O in BMB78 does not result from errors
within the analysis. Errors in the HF and OH abundances are discussed in detail in Cunha et al.
(2003) and Smith et al, (2003), respectively. Abundance uncertainties are expected to be
$\pm$0.15 dex for fluorine and $\pm$0.20 dex for oxygen. Since both HF and OH exhibit
similar sensitivities to changes in stellar parameters, their ratio is effectively
less sensitive to analysis uncertainties.  As BMB78 falls about 1.0 dex below the trend
defined by the other stars, analysis errors are unlikely to explain its low value of
F/O. Since the $^{19}$F-yield from SN II neutrinos is sensitive to
the metallicity of the supernova progenitor star, it is possible that BMB78 is a star that formed
from gas that was substantially enriched by ejecta from a metal-poor supernova.  Such a 
picture would indicate that metallicity in bulge stars proceeded in an inhomogeneous manner
at some level.  

This scenario can be tested, as $^{19}$F is not the only metallicity-dependent
element that has been studied in BMB78.  Sodium yields fom SN II are also metallicity dependent
and Na has been measured in BMB78 (Table 1).  Figure 3 displays results for sodium, where 
Na-to-O ratios
are plotted versus the oxygen abundance.  Field-star values of Na/O and A(O) are included
as the small blue open symbols. 
The solid curve contains the massive-star yields from WW95 convolved with a
Salpeter mass function.  Sodium yields are sensitive to stellar metallicity, with the Na-to-O
ratio increasing with increasing metallicity (taken here to be mapped by the oxygen
abundance), and the observed field star values track this curve quite well.  The bulge values
of Na/O and A(O) from Cunha \& Smith (2006) are shown as the large filled symbols with their
associated estimated errors: note that Na abundances for BMB78 and BMB289 are presented
here for the first time.  Additional bulge stars from Fulbright et al. (2007) and Lecureur et al.
(2007) are shown as the smaller filled symbols.  The agreement in the trend of Na/O with
A(O) is similar for all three bulge studies.

The sample of bulge red giants included in Figure 3 show some peculiarities
compared to the field stars.  
First there are the two Na-rich but O-poor giants from the Fulbright
et al. (2007) paper.  The pattern of Na/O and A(O) found in these two stars is very similar
to what is found in globular clusters and Fulbright et al. conclude that these two red giants are
actually members of the bulge globular cluster NGC6522 located in Baade's Window.  
All three bulge studies also contain a small number of stars that fall to the O-rich side of the distribution,
with lower Na-to-O ratios.  The star BMB78 is one of these examples, having a low Na abundance
when compared to its large oxygen abundance.  Since both F and Na have massive-star
yields that increase with metallicity, whereas O does not, the low values of F/O and Na/O in
this star can result from enrichment by a low-metallicity SN II.   Such a picture would suggest
that chemical evolution within the bulge population was not homogeneous.  The small number
of bulge stars that are found with lower values of Na/O may result from inhomogeneous
chemical evolution.

A picture of inhomogeneous chemical evolution can be checked for consistency as illustrated
in Figure 4, where the abundance ratios of F/Ti are plotted versus Na/Ti.  Titanium is chosen as
the fiducial element since there is evidence that oxygen yields are being altered at high
metallicity by metal-rich WR winds (McWilliam et al. 2007) and Ti typifies an $\alpha$-element
and thus serves as a monitor of  SN II enrichment.   In
this diagram the bulge stars fall along a sequence of increasing values of F/Ti with increasing
Na/Ti; the metallicity sensitive elements F and Na increase in lockstep and, in this case, BMB78
exhibits the lowest values of F/Ti and Na/Ti, which is consistent with processed gas from a
metal-poor SNII.

\section{Conclusions}

Fluorine abundances are measured for the first time in a sample of red-giants in the 
Galactic bulge. 
The fluorine abundances obtained generally define a steady increase in
F/O versus A(O), which is reminiscent of the disk results and can be explained
by production of $^{19}$F in a combination of AGB and WR stars. 
The most oxygen-rich target in this sample has a large fluorine abundance, but no 
accompanying s-process
enhancement, in contrast to the predictions for AGB nucleosynthesis by Goriely \& 
Mowlavi (2000).  The abundance pattern observed for this metal-rich bulge target favors $^{19}$F
production during the WR phase of evolution.
One oxygen-rich giant in this sample, however, fails to follow the disk trend and shows a fluorine
abundance, as well as sodium, that is more compatible with pollution from 
metal-poor SN II, where $^{19}$F is synthesized by neutrino nucleosynthesis. These results 
may indicate that there was inhomogeneous 
mixing in the gas that formed the Milky Way bulge during its phase of chemical enrichment.

\acknowledgements
We thank Andy McWilliam for kindly sending us bulge s-process results prior to publication
and the referee whose suggestions improved the paper.
This work is supported in part by the NSF 
(AST06-46790) and NASA (NAG5-9213).  
Based on observations obtained at the Gemini Observatory, which is operated by the
Assoc. of Univ. for Research in Astronomy Inc., under a cooperative agreement
with the NSF on behalf of the Gemini partnership: the NSF (United
States), the STFC (UK), 
the NRC (Canada), CONICYT (Chile), 
the ARC (Australia), CNPq (Brazil) and SECYT (Argentina).  
Based on observations obtained with the Phoenix
spectrograph, developed and operated by NOAO.

\clearpage

\begin{deluxetable}{lcccccc}
\tablecaption{Sample Stars and Derived Abundances}
\tablewidth{0pt}
\tablehead{
\colhead{Star} & \colhead{$T_{eff}$} & \colhead{Log g} & \colhead{$\xi$(km s$^{-1}$)} & \colhead{A(F)} &  \colhead{A(Na)} & \colhead{A(O)}}
\startdata
I-322      &  4250 & 1.5 & 2.0 & 4.50      & 6.13 & 8.60 \\
IV-003     &  4500 & 1.3 & 1.8 & ...       & 4.23 & 8.05 \\
IV-167     &  4375 & 2.5 & 2.2 & $<$6.10:  & 7.30 & 9.10 \\
IV-072     &  4400 & 2.4 & 2.2 & 5.60      & 7.35 & 9.20 \\
IV-329     &  4275 & 1.3 & 1.8 & 4.30      & 5.30 & 8.35 \\
BMB 78     &  3600 & 0.8 & 2.5 & 4.26      & 5.58 & 9.00 \\
BMB 289    &  3375 & 0.4 & 3.0 & 4.90      & 6.05 & 8.75 \\
\enddata
\end{deluxetable}

\clearpage

\begin{figure}
\includegraphics[scale=0.2,angle=270]{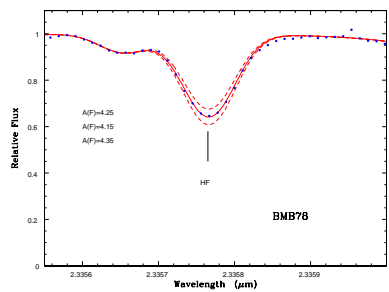}
\caption{\label{fig1} Observed (dotted line) and synthetic (solid and dashed lines) spectra
of the star BMB78 in the region of the HF line. 
The synthetic spectra were calculated for three fluorine abundances as specified in the figure.}
\end{figure}

\begin{figure}
\includegraphics[scale=0.3,angle=270]{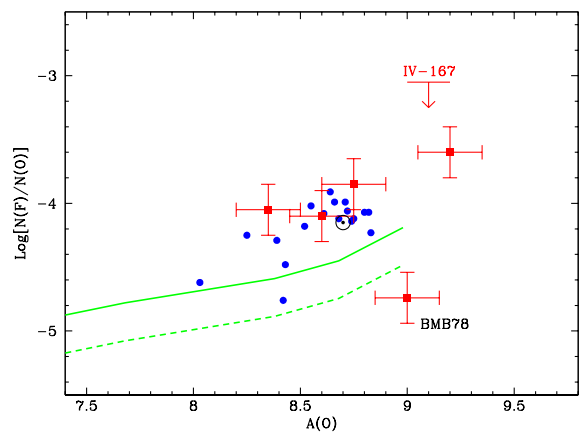}
\caption{\label{fig2} Ratios of F to O plotted versus the oxygen
abundance, A(O).  The values of F/O in 4 of the bulge stars track the trend defined for field stars,
with the O-rich star IV-072 extending the general field-star trend.  The bulge star BMB78 has a low
value of F/O for its O-abundance.  The solid curve illustrates model values of F/O versus A(O) for neutrino
nucleosynthesis from Woosley \& Weaver (1995), with the dashed curve representing 
a downward shift of the values of F/O as suggested by Heger et al. (2005).  }
\end{figure}

\begin{figure}
\includegraphics[scale=0.3,angle=270]{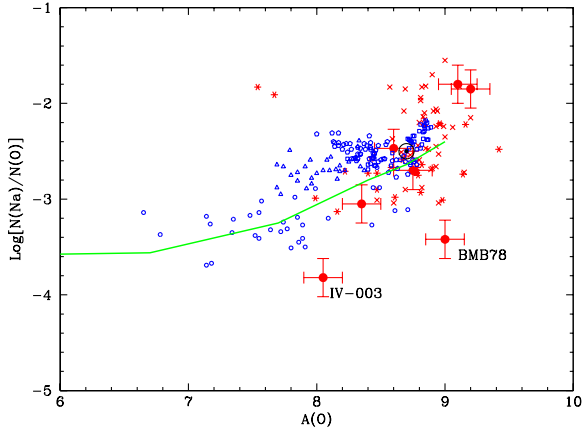}
\caption{\label{fig3} The behavior of Na/O versus O for bulge stars from this study
(red circles with errorbars), Fulbright et al. (2007 - small red asteriks) and
Lecureur (2007 - small red crosses). Galactic field star results are the small blue open symbols from
Nissen \& Schuster (1997), Fulbright (2002), Reddy et al. (2003), and Bensby et al. (2004).
The solid line represents yields from Woosley \& Weaver (1995) convolved with a standard
IMF. Note the position of BMB78, with a low ratio of Na/O at high metallicity; this
abundance pattern can result from enrichment by metal-poor SN II.
}
\end{figure}

\begin{figure}
\includegraphics[scale=0.3,angle=270]{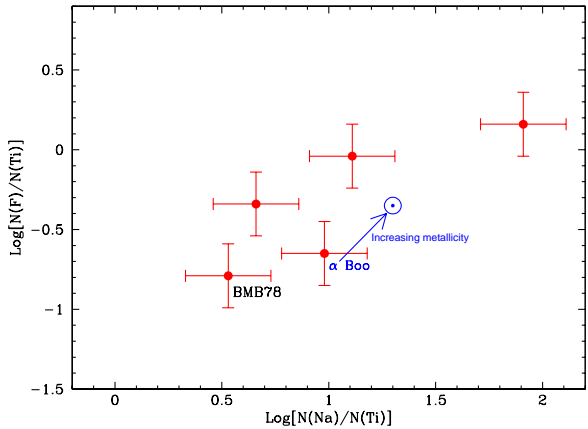}
\caption{\label{fig4} The run of fluorine over Titanium versus the abundances of
sodium over titanium for the bulge targets stars and the field star $\alpha$ Boo
(Cunha et al. 2003; Cunha \& Smith 2006). 
The position of the sun in this diagram is also shown for comparison. }
\end{figure}

\end{document}